\documentstyle[prl,twocolumn,aps,epsf] {revtex}
\begin{document}

\twocolumn [
\hsize\textwidth\columnwidth\hsize\csname@twocolumnfalse\endcsname

\title{Groundstates of $SU(2)$-Symmetric Confined Bose Gas:
Trap for a Schr\"odinger Cat}
\author{A.B. Kuklov$^1$ and B.V. Svistunov$^2$}
\address{$^1$ The College of Staten Island, CUNY, Staten Island, New York
10314}
\address{$^2$ Russian Research Center ``Kurchatov Institute", 123182 Moscow,
Russia}

\maketitle

\begin{abstract}
Conservation of the total isotopic spin ~$S$~of a two-component
Bose gas --- like~$^{87}$Rb --- has a dramatic impact on the structure of
the ground state. In the case when ~$S$~ is much smaller
than the total number of particles ~$N$, the condensation of each of the
two components occurs into at least {\it two} single-particle modes. The 
quantum
wavefunction of such a groundstate is a Schr\"odinger Cat --- a superposition
of the {\it phase separated} classical condensates, the most ``probable"
state in the superposition corresponding to the classical groundstate in the
sector of given $N$ and $S$.
\\


\end{abstract}
\vskip0.5 cm
]

Symmetry has profound implications
on the groundstate structure
of the degenerate superfluids and superconducting
systems.
A remarkable example to that
is superfluid ~$^3$He, which exhibits
a variety of textures \cite{HE3}.
These ground states
are essentially {\it classical} --- quantum fluctuations
play no role at large length scales.
Accordingly, the most important low energy physics
can be well described within the mean-field (MF)
approach relying on broken gauge symmetry.

Recent successes in trapping and manipulating
atomic Bose-Einstein condensate (BEC) \cite{BEC,MULT}
have initiated a great interest
to the physics beyond the MF in these systems.
A well-known example of the non-MF behavior is
the phase diffusion effect \cite{PD}.

A remarkable case of non-MF
ground state was found in the spin $s=1$
sodium vapor \cite{LAW,HO}. This state
is characterized
by anomalous quantum fluctuations of the components
with different projections of the total
spin.

In this Letter, we address the case of a two-component $SU(2)$-symmetric
Bose gas---like ~$^{87}$Rb \cite{RB}, that is, the case of
(pseudo-)spin-$1/2$
bosons. We show that, while sharing some common features with the
~$s=1$ bosons \cite{LAW,HO}, the $s=1/2$ system exhibits
unique properties \cite{note1}.
We pose a question of the formation of the ground state
in a naturally arising situation of ~$S\ll N$.
We find that the resulting ground state is built on at least {\it two}
macroscopically populated lowest
one-particle modes, and
is a {\it quantum superposition
of the phase separated BECs}, characterized by
large fluctuations
of the densities
of the components, the total density being fixed.

We start from the ~$SU(2)$-symmetric Hamiltonian

\begin{eqnarray}
\displaystyle
H=\int d{\bf x} \, [\Psi^{\dagger}_{\sigma}H_1
\Psi_{\sigma} + {g\over 2}\Psi^{\dagger}_{\sigma}
\Psi^{\dagger}_{\sigma'}
\Psi_{\sigma'}\Psi_{\sigma}] \; ,
\label{1}
\end{eqnarray}
\noindent
where the interaction constant ~$g=4\pi a/m$~
is represented (in units ~$\hbar =1$~) by the scattering length ~$a$~ and
mass ~$m$;
the single particle Hamiltonian ~$H_1$~
consists of the kinetic energy and the trapping potential,
~$H_1=-(1/2m){\bf \nabla}^2
+U({\bf x})$; the Bose-field operators ~$\Psi_{\sigma}$
have two components ~$\sigma =\uparrow , \downarrow $
in pseudo-spin notations, and
the summation is performed over repeated
pseudo-spin (Greek) indices.
Due to its symmetry, this Hamiltonian conserves
the total isospin ~${\bf S} = (S_x,\,S_y,\,S_z)$~ given as (~$S_{\pm}=S_x\pm
iS_y$ ,~ $S_+ \equiv S_-^{\dagger}$)

\begin{eqnarray}
S_z&=&{1\over 2}\int d{\bf
x}\, (\Psi^{\dagger}_{\uparrow}\Psi_{\uparrow}-
\Psi^{\dagger}_{\downarrow}\Psi_{\downarrow})\, , \;\;
S_-=\int d{\bf x}\, \Psi^{\dagger}_{\downarrow}\Psi_{\uparrow}\; .
\label{12}
\end{eqnarray}
\noindent
These operators obey the standard ~$SU(2)$~ algebra
~$ [S_z,S_+]=S_+,\,\,\, [S_z,S_-]=-S_-,\,\,\,\,
[S_+,S_-]=2S_z$.

The absolute minimum of energy is achieved when all
atoms condense to a lowest one-particle (Hartree)
state ~$\varphi_0$. Such a state
corresponds to the maximum possible
value of the isospin
~$S_{\rm max}=N/2=(N_{\uparrow}+N_{\downarrow})/2$, where
~$ N_{\uparrow},\, N_{\downarrow}$~ stand
for the total numbers of atoms of each
component. Thus, it follows that a very
nontrivial situation
will occur when, while conserving ~$S$, the
condensation proceeds from the thermal
cloud which contains almost the same
amount of different components
~$N_{\sigma}$~
with {\it no inter-component correlations}.
Indeed, the ensemble mean
of the total isospin (\ref{12})
in Boltzmann gas
is ~$\langle {\bf S}\rangle =0$,
and the mean of ~$
{\bf S}^2=S^2_z + \left( S_+ S_ -
+ S_- S_ +\right)/2$~ yields
~$\langle {\bf S}^2\rangle \sim N$,
as long as
~$\langle
\Psi^{\dagger}_{\uparrow}({\bf x}) \Psi_{\downarrow}({\bf x}')
\rangle =0 $~ and the diagonal
correlators
decay on thermal length.
Similarly, ~$\langle {\bf S}^2\rangle \sim N$,
when two spatially
separated degenerate components
are mixed together (see below).
Thus, the resulting ensemble of the
ground states is characterized
by typical values ~$0\leq S\sim \sqrt{N}$.

Since condensation of both components into one single-particle
state $\varphi_0$
is incompatible with $S \ll N$,
the resulting many-body state is
characterized by macroscopic occupation
of at least {\it two} lowest one-particle states.
This contrasts the cases ~$s=1/2$~ and ~$s=1$ \cite{LAW,HO}.

We analyze the simplest case of the
strong quasi-1D trap, where the energy
difference ~$\varepsilon =\varepsilon_1-\varepsilon_0$~
between the lowest single particle levels ~$\varphi_{1,0}$,
respectively, in the trapping potential
~$U$~ is much larger than a typical
interaction energy per particle ~$\mu$~ (chemical
potential at ~$T=0$). The inequality ~$\mu/\varepsilon \ll 1$~
allows one to essentially simplify the description by
introducing a projected 
Hamiltonian acting in a truncated Hilbert space
containing only two single-particle modes, $\varphi_0$ and $\varphi_1$:
\begin{eqnarray}
\displaystyle
\Psi_{\sigma}=\sum_{b=0,1} a_{b\sigma}\varphi_b
\label{two}
\end{eqnarray}
\noindent
where $a_{b\sigma}$
annihilates a particle at the level $b$ in the spin state $\sigma$.
With the same accuracy, one may neglect in the projected Hamiltonian
all the terms that do not conserve the total occupations of each of the
two single-particle levels. These terms involve a large energy difference
$\sim \varepsilon \gg \mu$, and, thus, result only in
higher-order---in parameter $\mu/\varepsilon$---perturbative
renormalization of the parameters of the effective 
(two-modes \& two-colors) Hamiltonian

\begin{eqnarray}
\displaystyle
H'&=&\sum_{a=0,1}\varepsilon_aN_a +
\nonumber
\\
&+&{1\over 2}\sum_{a,b=0,1}I_{ab}N_a(N_b-\delta_{a,b}) + 2I_{01}{\bf S}_0
{\bf S}_1,
\label{3}
\end{eqnarray}
\noindent
where ~$I_{ab}=g\int \, d{\bf x} \, |\varphi_a|^2|\varphi_b|^2$~
and
~$N_b=\sum_{\sigma}N_{b\sigma},\,\,
S_{bz}=( N_{b\uparrow} - N_{b\downarrow})/2,\,\,
S_{bx}=( a^{\dagger}_{b\uparrow} a_{b\downarrow}
+ a^{\dagger}_{b\downarrow} a_{b\uparrow})/2,\,\,
S_{by}=( a^{\dagger}_{b\uparrow} a_{b\downarrow}
- a^{\dagger}_{b\downarrow} a_{b\uparrow})/2i$;
the operators ~$N_{a\sigma}= a^{\dagger}_{b\sigma} a_{b\sigma}$~
and ~${\bf S}_a
=(S_{ax}, S_{ay}, S_{az})$~ represent the
numbers of atoms and the components of the pseudo-spin
on the $a$th level, respectively. The quantities $N_a$
and ${\bf S}_a^2$ are related to each other by the identity
${\bf S}_a^2= N_a(1 +N_a/2)/2$, implying that $S_a \equiv N_a/2$.

A standard choice of eigen numbers for the Hamiltonian ({\ref{3}) comes
from the theory of two spins: the total spin $S$,
its $z$-projection, $M$, and the spins $S_0$ and $S_1$;
with the constraints $|S_0-S_1| \leq S \leq S_0+S_1$, $-S \leq M \leq S$.
In this nomenclature, the spectrum of (\ref{3}) is
\begin{eqnarray}
\displaystyle
E_{S M S_0 S_1}&=&2 \varepsilon_0 (S_0+S_1) + 2\varepsilon S_1 +
\sum_{a,b}I_{ab}S_a(2S_b-\delta_{a,b})
\nonumber
\\
+I_{01}[S(S&+&1) - (S_0+1)S_0 -(S_1+1)S_1] \, .
\label{5}
\end{eqnarray}
The distribution over the eigennumbers ~$S_0,\, S_1, \, S,\, M$,
which will be formed in the end of the cooling processes,
depends essentially on the following:
While $S$, $M$ and
$N$ {\it are} the eigennumbers of ~$H$, eq.(\ref{1}),
the spins ~$S_0,\, S_1$~ {\it are not} the eigennumbers of the
original Hamiltonian (\ref{1}).
Thus, the formation of particular low
energy states, amenable to treatment by the reduced
Hamiltonian (\ref{3}) where ~$S_{0,1}$~ are now
the quantum numbers, will be controlled by the
microcanonical distribution with fixed values
of ~$S,\,M,\,N$.
For our purposes,
we do not need to adopt any particular distribution for $S$,
$M$ and $N$. We assume that their values are obtained (after
the cooling is over) by
a non-destructive measurement. Thus,
the only remaining degree of freedom,
within the chosen microcanonical
ensemble,
is the integer $0 \leq K \leq 2 S$, that
parameterizes $S_1=(K-S + N/2)/2$ and $ S_0=(-K+S + N/2)/2$~
in eq.(\ref{5}).
The corresponding Gibbs distribution over $K$
with the temperature created by the cooling follows
from the spectrum eq.(\ref{5}), which now takes
a form
~$E_{SMN}(K)=(\varepsilon_0+ \varepsilon /2)N+
\varepsilon (K-S) +o(\mu/\varepsilon)$~
in the  main ~$\varepsilon \gg \mu$~ limit.
It indicates that
the ground state for given ~$S,M,N$~ corresponds
to ~$K=0$. The next excited ~$K=1$~ state within the microcanonical
ensemble costs energy ~$\varepsilon$.
Thus, once temperature ~$T< \varepsilon$,
the microcanonical ensemble for given ~$S,M,N$~
essentially consists of only one state---the
ground state ~$K=0$, given as
\begin{eqnarray}
\displaystyle
|SMN\rangle&=&Z_{SMN}\left(a^{\dagger}_{0\uparrow}\right)^{S+M}
\left(a^{\dagger}_{0\downarrow}\right)^{S-M}
\left(R^{\dagger}\right)^{N/2-S}|0\rangle,
\label{7}
\\
R&=&a_{0\uparrow} a_{1\downarrow}- a_{1\uparrow} a_{0\downarrow},
\nonumber
\end{eqnarray}
\noindent
where $S_0=(S+N/2)/2$ and $S_1=(-S+N/2)/2$;
$ Z_{SMN}$~ is the normalization coefficient, and $|0\rangle$ is the
vacuum.
These states realize the irreducible representation
of the isotopic $SU(2)$ group. For $S=0,M=0$~ (~$N$~ even),
the state (\ref{7}) is the $SU(2)$-singlet.

The state (\ref{7}) exhibits quantum fluctuations of individual
values of $N_{a\sigma}$,
preserving, however, their three linear combinations:
~$ N_{0\uparrow} + N_{0\downarrow} =S+N/2,\,\,
N_{1\uparrow} + N_{1\downarrow}=-S +N/2,\,\,
N_{1\uparrow} + N_{0\uparrow}
- N_{1\downarrow} - N_{0\downarrow}=2M$.
Thus, there is only
{\it one} degree
of freedom, which we choose to be $N_{0\uparrow}$.
Most transparently, these fluctuations are described
in the basis of the Fock states
~$|N_{0\uparrow},N_{0\downarrow}, N_{1\uparrow},
N_{1\downarrow}\rangle$~, which are the
{\it fragmented condensates} \cite{FRAG}. Expanding (\ref{7}) over the Fock
basis,
we get the  probability to find a given value of $N_{0\uparrow}$:
\begin{eqnarray}
\displaystyle
P_{SMN}(N_{0\uparrow})=
{Z^2_{SMN} N_{0\uparrow}!
(N/2+S- N_{0\uparrow})!\over (N_{0\uparrow}-S-M)!
(N/2+M- N_{0\uparrow})!} \; .
\label{8}
\end{eqnarray}
\noindent
To estimate the strength of the quantum fluctuations,
consider $\overline{\delta N_{0\uparrow}^2}$, where for any $Q$:
$\overline{Q}=\langle SMN|Q|SMN \rangle$ and
~$\overline{\delta Q^2}
=\langle SMN|(Q-\overline{Q}
)^2|SMN\rangle$. At $S=0$, we have
~$P_{SMN}(N_{0\uparrow}) =2/N
$~ over the whole range of ~$ 0\leq N_{0\uparrow}
\leq N/2$; and
~$\overline{\delta N^2_{0\uparrow}}=
N^2/12$. For ~$1\ll S\ll N,\,\, S-|M| \gg 1$, the Gaussian
approximation of (\ref{8}) gives

\begin{eqnarray}
\displaystyle
\overline{N}_{0\uparrow}&\approx& (S+M)(S+N/2) / 2S \; ,
\label{9_0}
\\
\overline{\delta
N_{0\uparrow}^2}
&\approx &  (S^2-M^2)((N/2)^2-S^2) / 8S^3 .
\label{9}
\end{eqnarray}
\noindent
For typical values ~$S\sim \sqrt{N}$,
we obtain
~$\overline{\delta N^2_{0\uparrow}}\approx N^2/( 8S)
\sim N^{3/2}$. In a direct analogy with the $s=1$ case \cite{LAW},
these anomalous fluctuations are consequence of the conservation
of the total (iso)spin \cite{HO}.

The 
 state (\ref{7}) is
a quantum superposition --- a Schr\"odinger
Cat --- of the {\it phase separated} ``classical" BEC's, each  BEC
being characterized by
well-defined spatial density distribution of the
spin-up and spin-down components. The phase separation is
due to the involvement of 
two single-particle modes with
different occupations of the
spin-up/down components.

Measurement of ~$N_{0\uparrow}$~
produces a {\it collapse} of the Cat.
The resulting states
essentially depend on how strong the interaction between the measuring
device and the system is.
The most ``accurate" measurement of $N_{0\uparrow}$, with the absolute
error less than one particle, would collapse the Cat into
a Fock state $|N_{0\uparrow},N_{0\downarrow}, N_{1\uparrow},
N_{1\downarrow}\rangle$, with the probability (\ref{8}) to obtain given
$N_{0\uparrow}$. Such a measurement, however, implies that the interaction
of the system with the apparatus is much stronger than that due to the
Hamiltonian (\ref{3}). In this sense, such a measurement is destructive
not only to the supposed-to-be-fragile Quantum Cat,
but also to its classical counterparts.

More interesting is a less destructive measurement,
which collapses Quantum Cat into some ``classical"
(in a particular sense) states. These classical
states are the phase-coherent BEC states,
which can be interpreted
in terms of the classical vectors  ~${\bf S}_0,\, {\bf S}_1$
interacting
anti-ferromagnetically by the term ~$2I_{01}
{\bf S}_0 \cdot {\bf S}_1= I_{01}[
S^2 - S_1^2 -S_0^2],\,\, I_{01}>0$.
Comparing the classical
energy with the term
in the square brackets in the full (quantum) expression (\ref{5}),
we see that the creation of such classical states
costs only the energy ~$\delta E_C\approx
I_{01}[S_0 + S_1] \sim I_{01}N \sim \mu$.
Here we took into account that ~$S_0\sim S_1 \sim N$~
for ~$S \ll N$.

The above-discussed classical
BEC states are readily obtained within the Gross-Pitaevskii approach.
As an instructive example, let us construct the classical groundstates.
To this end we replace the bosonic fields operators
by the $c$-fields. Accordingly, in eqs.~(\ref{two},\ref{3}),
we replace the ~$a$-operators by
the $c$-number amplitudes, and find the minimum of energy at
given $N,\, S,\, M$. This yields
~$a_{0\uparrow}=
\sqrt{S+ N/2}\, \cos(\beta/2)\exp(i(\alpha_0+\gamma)/2),\,\,
a_{0\downarrow}=-
\sqrt{S+ N/2}\, \sin(\beta/2)\exp(i(\alpha_0-\gamma)/2),\,\,
a_{1\uparrow}=i
\sqrt{-S+ N/2}\, \sin(\beta/2)\exp(i(\alpha_1+\gamma)/2),\,\,
a_{1\downarrow}=i
\sqrt{-S+ N/2}\, \cos(\beta/2)\exp(i(\alpha_1-\gamma)/2)$.
Here ~$\cos \beta =M/S,\,\, (\alpha_1-\alpha_0)/2=
-\varepsilon (1 +o(\mu /\varepsilon))t$;
~$\alpha_0$~ is the global phase, and ~$\gamma$~
determines the $x,y$-projections of the
total spin (~$S_x=-S\sin \beta \cos \gamma$).
Comparison of
these with eq.(\ref{9_0}) shows
that the classical occupation numbers coincide with
the expectations for the quantum ones.

The classical fields are
non-stationary,
which is a consequence of the energy difference between the
states $\varphi_0$ and $\varphi_1$. Hence, wherever $\varphi_0({\bf x})$
and $\varphi_1({\bf x})$ have substantial overlapping, the spatial
densities of the two components experience strong fluctuations, the total
density  remaining constant. These fluctuations are of
purely classical nature and have nothing to do with the above-discussed
quantum fluctuations of occupation numbers! It is reasonable
to reveal these classical fluctuations directly in the quantum
wavefunction (\ref{7}). To this end we calculate fluctuations
of the difference of the component densities, ~$\eta ({\bf x})
= \Psi^{\dagger}_{\uparrow}({\bf x})
\Psi_{\uparrow}({\bf x}) -
\Psi^{\dagger}_{\downarrow}({\bf x})
\Psi_{\downarrow}({\bf x})$~ [Note that
~$\overline{\eta}\approx 0$~
for ~$|M|\ll S$, that is the state (\ref{7})
{\it does not} exhibit any phase separation
{\it on average}]. For the fluctuations of
$\eta$ we get
\begin{eqnarray}
\displaystyle
\overline{\delta \eta^2}&=&
4[\varphi_0^4 +\varphi_1^4 -4\varphi^2_0
\varphi_1^2]\overline{\delta
N_{0\uparrow}^2} + 2\varphi_0^2\varphi_1^2
[(N /4) ( N +4 )   -
\nonumber
\\
&-&S(S+1) + 4\overline{S_{0z}}(
M-\overline{S_{0z}} ) ]\; .
\label{S}
\end{eqnarray}
\noindent
The two terms in (\ref{S}) correspond to the quantum
($\sim \overline{\delta
N_{0\uparrow}^2}$) and classical
($\sim N^2$~ for ~$S\ll N$)
fluctuations of $\eta$, respectively. Though the second term is dominant
in a general case, it vanishes in the trap center, where $\varphi_1=0$,
and the fluctuations are of purely quantum nature.

To propose a particular set of measurements that could reveal the
statistics (\ref{8}), we note that the quantities
$N_0=\sum_{\sigma}N_{0\sigma}= N/2+S,\,\,
N_1=\sum_{\sigma}N_{1\sigma}= N/2-S,\,\,
N_{\uparrow} =N_{0\uparrow}+ N_{1\uparrow}= N/2+M,\,\,
N_{\downarrow}=N_{0\downarrow}
+ N_{1\downarrow}= N/2-M$~
{\it do not} fluctuate.
Thus, the values ~$S,M,N$~ can be extracted, in effect,
{\it without} destroying the wave function.
Once the set of values ~$S,M,N$ is fixed,
one can measure $N_{0\uparrow}$, by, e.g., observing
the density of the spin-up component in the center of the
trap by the spatially differentiated selective imaging \cite{IMAGE}.
In the trap center $\varphi_1=0$, therefore the component
densities are unambiguously related to the occupations of
the state $\varphi_0$ \cite{note3}.
Actual measurements always have
some uncertainties ~$\Delta M,\,
\Delta S,\, \Delta N$~
in determining ~$M,\, S,\, N$, respectively.
These may wash out the quantum fluctuations.
First,
we note that the mean value ~$\overline{N}_{0\uparrow}$~
in eq.(\ref{9_0})
is exactly the value given within the classical
geometrical picture. Thus,
~$\langle (\Delta \overline{N}_{0\uparrow})^2\rangle$,
where ~$ \langle ... \rangle$~
denotes the statistical mean,
must be significantly smaller
than the quantum fluctuation ~$
\overline{\delta
N_{0\uparrow}^2}
$, eq.(\ref{9}).
Practically, for ~$\langle (\Delta N)^2\rangle \leq N,\,\,
\langle S \rangle \sim \sqrt{N}$,
we find that, as long as ~$\langle (\Delta M)^2\rangle
\ll \langle S\rangle/2$, the
quantum fluctuations can be well
distinguished.

We consider an intriguing alternative to the
direct cooling of the two-component system in order to
obtain the state (\ref{7}). Namely, the {\it merging} of
two spatially separated condensed species from
two identical wells with equal numbers of
oppositely "polarized" particles in each of them.
At ~$T\ll \varepsilon$~ 
(which means, for our purposes, practically zero),
and in the absence of tunneling
between the wells, the system is in its lowest energy
state. We note further that our state is
$(N/2+1)$-fold degenerate within the given sector of $N$ and $M=0$,
the other $N/2$ groundstates being obtained by
exchanging spin-up--spin-down pairs between the wells.
The key point is that the initial state is
a superposition of the eigenstates of the total
pseudo-spin operator with the typical values ~$0\leq S \lesssim \sqrt{N}$,
and the degeneracy is exhausted by the isospin
index (which, at $M=0$, has exactly $N/2+1$ values, from $0$ to $N/2$).
Thus, each of the $S$-eigenstates in our superposition is {\it not
degenerate}. Accordingly, when
tunneling is slowly turned on and, finally,
the inter-well barrier is adiabatically removed \cite{NOTE4},
the system remains in the superposition
of the ground $S$-eigenstates, which at $\varepsilon/\mu \gg 1$ 
take on the form (\ref{7}) (~$M=0$).
Projection on the particular $S$  can be done by, say,
measuring total density in the trap center \cite{note3}.

Other (more exotic) options for the direct Cat-state engineering
rely on merging more than two BEC clouds
in the presence of the controlled $SU(2)$ symmetry
violating potential. We will analyze them elsewhere.

Let us discuss the effect
of the {\it symmetry breaking static perturbation},
which was shown to destroy the
quantum fluctuating state in the
case ~$s=1$~  in the thermodynamical
limit \cite{HO}.
A similar situation occurs
in the case of the pseudo-spin ~$s=1/2$~ bosons.
A role of the perturbation
is played by the gradient ~${\bf B}'$~ of the
pseudo-magnetic field ~$
{\bf B}'=(0,0,B')$, consisting of
the difference ~$U'=U_{\uparrow} -
U_{\downarrow}$~ of the trapping potentials
for the corresponding components and of
the interaction term ~$\tilde{H}=
(g'/2)\int\, d{\bf x} \,[\Psi^{\dagger}_{\uparrow}
\Psi^{\dagger}_{\uparrow}\Psi_{\uparrow}
\Psi_{\uparrow}
-\Psi^{\dagger}_{\downarrow}
\Psi^{\dagger}_{\downarrow}\Psi_{\downarrow}
\Psi_{\downarrow}]$, where ~$g'=4\pi a'/m$~
and ~$|a'|\ll a$~ in ~$^{87}$Rb \cite{RB}.
Within the model,
we find the corresponding term

\begin{eqnarray}
\displaystyle
H_B&=&-B'(S_{0z} - S_{1z}),
\nonumber
\\
B'&=&\int d{\bf x}\,( U'({\bf x})+{g'\over 2}\rho
({\bf x}))(\varphi^2_1-
\varphi^2_0),
\label{90}
\end{eqnarray}
\noindent
to be added to ~$H'$, eq.(\ref{3}),
where we take
into account ~$N_0 \approx N_1\approx N/2$,
and
~$\rho({\bf x}) =N(\varphi^2_0 +\varphi^2_1)/2$.

The term ~$H_B$~
does not commute
with the total spin, which implies that it
{\it does not} favor
quantum fluctuations.
It decreases energy of the classical (not
fluctuating) state
~$|BEC\rangle=(a^{\dagger}_{0\uparrow} a^{\dagger}_{1
\downarrow})^{N/2}|0\rangle$~ by
~$\delta E_{BEC}
\approx -\langle BEC|H_B|BEC \rangle
\approx B'N$,~ (~$B'>0$). At the same
time, the classical
states loose energy with respect
to the lowest eigenstate of ~$H'$~, eq.(\ref{3}),
by the chemical potential ~$\mu$. Accordingly,
in order to insure the stability of
the Schr\"odinger Cat state (\ref{7}),
the condition
~$\mu \geq |B'|N$~
should hold. This condition cannot
be obviously satisfied in the
thermodynamical limit, and
should be considered in mesoscopic
situations only. We note, however, that it
can be well controlled within the current
experimental capabilities.
Taking the effective
potential breaking the symmetry
as the oscillator
potential ~$\delta U=m\nu^2x^2/2$, with
some frequency ~$\nu$ \cite{JILA2},
we find
~$B'=\nu^2/(2\varepsilon)$ in eq.(\ref{90}),
where we have employed  the lowest
eigenstates ~$\varphi_{0,1}$,
of the 1D oscillator with frequency
~$\varepsilon$.
Accordingly,
~$\nu$~ should obey
~$ \nu /\varepsilon
\leq \sqrt{\mu /(\varepsilon N)}$.
For ~$N\leq 10^4$, ~$\mu \leq \omega_0$,
this yields
~$\nu/\varepsilon \leq 10^{-2}$, which is
well within the accuracy of
the experiment \cite{JILA2}.
In fact, the control of the
value ~$B'$~ together
with the external rf-pulses \cite{RB,IMAGE}
provides
a tool for manipulating the parameters
of the Cat state.

Another effect of the symmetry breaking,
existing even for ~$B'=0$,
is opening the channel for irreversible
transitions ~$1\to 0$. While
bringing the system to the absolute ground
state (~$S=N/2$), it is, however,
of a {\it second} order in ~$g'$, and
needs the normal component to be present
to proceed at noticeable rate.
In $^{87}$Rb \cite{RB}, ~$g/g'\approx 30$.
Thus, a time scale for a such relaxation
is a factor of ~$10^3$~ longer than
relaxation time $\sim 1/g^2$
of any lowest energy excitation
at considered low temperatures and densities,
which makes it irrelevant for all practical
purposes, if compared with the time
of the Cat-state formation ~$\sim 1/\mu \sim 1/g$~
 \cite{NOTE4}.

Let us consider stability of the
above analysis with respect to external 
{\it symmetry breaking temporal noise}. 
For example,
the trapping potential may be slightly different
for the components and can fluctuate in time.
These fluctuations may produce two qualitatively
different effects: 1) heating of the cloud at some
rate ~$\tau^{-1}_h$, and 2) disruption of the
coherence at a rate ~$\tau^{-1}_{irr}$, which,
generally speaking, is different from ~$\tau^{-1}_h$.

A most detrimental for the Cat situation may occur
when the noise, while producing no significant
heating, disrupts the anomalous quantum correlations
~$A(t)=\overline{\delta N^2_{0\uparrow}}$,
as given by eq.(\ref{9}), at times ~$\tau_{irr}\ll
\tau_h$. 
As discussed in ref.\cite{DUBNA}, this occurs
when the noise correlation length is larger
than a typical thermal length of the cloud. 
Accordingly, if
the noise is produced by small correlation
length factors, such as, e.g., collisions with
the background gas, the decoherence {\it is not}
faster than the heating.

In order to investigate the case
under consideration,
we note that $A(t)$ can naturally be represented
in terms of the 
one-particle 
~$\rho^{(1)}({\bf x}, {\bf x}',t)=
\langle \Psi^{\dagger}({\bf x},t) \Psi({\bf x}',t)\rangle$~ (OPDM)
and two-particle ~$\rho^{(2)}({\bf x}_1, {\bf x}_2,
{\bf x}'_2, {\bf x}'_1, t)=
\langle \Psi^{\dagger}({\bf x}_1,t) \Psi^{\dagger}({\bf x}_2,t)
 \Psi({\bf x}'_2,t) \Psi({\bf x}'_1,t)\rangle$~ (TPDM) density
matrices, respectively, where the averaging
is performed over the initial state, which
is taken to be the Cat-state (\ref{7}),
 and over the noise. 
Accordingly, the equations for
them should be analyzed and the rates ~$\tau^{-1}_{irr}$,
 ~$\tau^{-1}_h$~ evaluated and, then, compared with each other.
In the case of the white noise, this can be done, practically,
exactly with the help
of the Furutzu-Novikov theorem as discussed in ref.\cite{DUBNA}. 

The heating rate is defined as the inverse time ~$\tau_h$~
required to increase the total
kinetic energy per particle
~$ K(t)=-(1/m)\nabla^2_{{\bf x}={\bf x}'}
\rho^{(1)}({\bf x}, {\bf x}',t)/N \sim t$~ by ~$\varepsilon$.
The decoherence rate ~$\tau_{irr}^{-1}$~
 is defined as the inverse of a typical {\it shortest } time of the
decay of the elements of the OPDM and the TPDM.  
All rates can be evaluated with respect to the OPDM and the TPDM
solutions
in the limit ~$t\to 0$. 

It is important to note that, 
while the heating is most effectively produced by a short
range noise, the coherence is destroyed by the long
range noise due to the
long range temporal
fluctuations of the trapping potential ~$U'({\bf x},t)$.
These fluctuations can be produced by 
thermal or any other noise of electric currents ~$I'(t)$~
in the magnetic coils. For the purpose
of estimating, we assume that 

\begin{eqnarray}
\displaystyle
U'=U({\bf x})I'(t)/I,\,\, \langle I'(t)I'(t')\rangle=
C^2\delta (t-t'),
\label{U'}
\end{eqnarray}
\noindent
where ~$U({\bf x})$~ is some typical trapping
potential, which does not fluctuate; ~$I$~ stands 
for the static current; we assumed the white noise
structure of the current fluctuations, which
is characterized by the constant ~$C^2$. In the case
of large temperatures of the coils ~$T_C$, this constant
is given by the fluctuation-dissipation theorem as
~$C^2=2T_C/R$, where ~$R$~ stands for the resistance.
Proceeding
as described above and in ref.\cite{DUBNA} and taking 
into account eq.(\ref{U'}), 
the heating and the decoherence rates
can be found as 

\begin{eqnarray}
\displaystyle
\tau^{-1}_h\approx \overline{(\nabla U({\bf x}))^2}
C^2/(m\varepsilon I^2),\,\,
\tau^{-1}_{irr}
\approx \overline{U^2({\bf x})}
C^2/I^2,
\label{IRR}
\end{eqnarray}
\noindent
respectively, 
where the line      
stands for a typical average over the cloud volume,
and the numerical coefficients
of order 1-10 (depending
on the order of the correlator) are omitted.

We
estimate ~$\overline{U^2({\bf x})} \approx \varepsilon^2$~ 
and take into account that typical
kinetic energy is of the order of the potential
energy, that is,
~$\overline{(\nabla U({\bf x}))^2}/m \approx \varepsilon^3$.
For the thermal noise,
 where ~$C^2=2T_C/R$, ~$T_C=300K$ (room temperature), 
$\varepsilon /\hbar \approx 10^3$s$^{-1}$,
we find
~$\tau^{-1}_{irr}\approx \varepsilon^2 k_BT/(\hbar^2 RI^2)\sim 10^{-15}$s$^{-1}$,
if the coil power ~$RI^2\sim 1$W
(we restored the standard units). Accordingly, the thermal
noise of the coil currents can be ignored for all practical
purposes. 

In a general case of a non-thermal noise characterized by
some relaxation time ~$\tau_C$~ and the current
fluctuation magnitude ~$\langle I^{'2}\rangle$~, so that 
$\langle I'(t)I'(t')\rangle=\langle I^{'2}\rangle\exp (-|t-t'|/\tau_C)$,
the white noise result \cite{DUBNA} and the estimate (\ref{IRR})
can still be employed as long as ~$\tau_C \ll \tau_{irr}$. We, then,
take ~$C^2\approx \tau_C\langle I^{'2}\rangle$~, and obtain
~$\tau^{-1}_{irr}\approx (\varepsilon /\hbar)^2 \tau_C\langle I^{'2}\rangle /I^2$.
This implies that, for , e.g., 
~$\tau_C=10^{-6}$s, $\varepsilon/\hbar \approx 10^3$s$^{-1}$, the
relative 
fluctuations ~$\langle I^{'2}\rangle /I^2$~ can be as large as 10\% and yet  
~$\tau_{irr}> 100$s, which should be compared
with the time of the Cat-state formation \cite{NOTE4}.
As discussed above, this time can be as short as
~$\sim 1/\mu \sim 10^{-2} - 10^{-3}$s.

The decoherence due to losses can similarly be analyzed within
the framework of equations for the OPDM and TPDM. Regardless of
the nature of the interaction vertex responsible for the losses,
these equations can be obtained as an expansion with respect
to the vertex, and, then, the rate ~$\tau^{-1}_{irr}$~ compared
with the corresponding contributions to the
rate of losses ~$\tau^{-1}_{loss}$, defined as the time derivative
of ~$\int d{\bf x}\rho^{(1)}({\bf x}, {\bf x},t)$~ per one particle.
Performing this general analysis, we find that ~$\tau^{-1}_{irr}$~
and ~$\tau^{-1}_{loss}$~ are of the same order.
Thus, once the losses are kept low 
during the time of the experiment (the scale of which is
determined by the adiabaticity requirement
\cite{NOTE4}), the decoherence can be ignored.

{\it In summary},
we have discussed the role of the isotopic
$SU(2)$ symmetry in the formation of the ground
states in a confined mesoscopic two-component
Bose gas. The conservation of the total
isotopic spin results in condensation
of a thermal cloud, consisting of the two
uncorrelated components, on {\it two}
lowest single particle states in the trap.
The resulting ground state is essentially
non-classical Schr\"odinger Cat
state characterized by
anomalously large non-uniform fluctuations of the
density of each component about its classical
expectation. The symmetry
breaking term may destabilize this
state into the phase separation, with
both components becoming the classical BEC.
This process, however, can be well controlled
experimentally.
The realization of the fluctuating many body
state provides unique opportunity for
observing the effect of measurement induced
collapse of the Schr\"odinger Cat to
classical states. The conditions required
for distinguishing the quantum fluctuation
from the statistical noise are delineated.

The authors are grateful to
N.P.~Bigelow, M.J.~Holland,
A.E.~Meyerovich, N.V.~Prokof'ev,
and Li You
for their interest to this work and useful
discussions. ABK was supported by CUNY grant PSC-63499-0032.
BVS acknowledges a support from Russian Foundation
for Basic Research under from the
Netherlands Organization
for Scientific Research (NWO).


\begin{references}
\bibitem{note1} A crutial circumstance that distinguishes $s=1/2$ bosons
from
the $s=1$ ones is the conservation of the number of particles of each spin
component.
Hence, apart from a special symmetry of the Hamiltonian, $s=1/2$ bosons can
be
considered as a mixture of two components, each component possessing a
well-defined number of particles.
\bibitem{note3}
For example, measurement of the total density
in the center gives ~$(N/2 +S)\varphi_0^2(0)$,
which yields ~$S$. This should be contrasted
with the density
~$N\varphi_0^2(0)$~ in the absolute
ground state ~$S=N/2$.
\bibitem{NOTE4}
Initially, within the given $S$-sector, the lowest excited
state costs energy ~$\sim \mu$. Thus, the adiabaticity
time scale must be ~$ \gg 1/\mu$.
\end{references}
\end{document}